\documentclass[12pt]{article}
\usepackage{amssymb}
\usepackage{color}
\usepackage{graphicx}

\newcommand{\A}{{\mathfrak A}}

\newcommand{\R}{{\cal R}}

\newcommand{\mc}{\mathcal}

\newcommand{\be}{\begin{equation}}
\newcommand{\en}{\end{equation}}
\newcommand{\bea}{\begin{eqnarray}}
\newcommand{\ena}{\end{eqnarray}}

\newcommand{\1}{1 \!\! 1}
\newcommand{\ST}{\mc S}

\newcommand{\Hil}{\mc H}

\catcode `\@=11 \@addtoreset{equation}{section}
\def\theequation{\arabic{section}.\arabic{equation}}
\catcode `\@=12

\begin{document}

\begin{center}
{\Large \bf Simplified stock markets described by number operators}   \vspace{2cm}\\

{\large F. Bagarello}
\vspace{3mm}\\
  Dipartimento di Metodi e Modelli Matematici,
Facolt\`a di Ingegneria,\\ Universit\`a di Palermo, Viale delle Scienze, I - 90128  Palermo, Italy\\
E-mail: bagarell@unipa.it\\home page:
www.unipa.it$\backslash$\~\,bagarell
\vspace{2mm}\\
\end{center}

\vspace*{2cm}

\begin{abstract}
\noindent In this paper we continue our systematic analysis of the
operatorial approach  previously proposed in an economical context
and we discuss a {\em mixed} toy model of  a simplified stock market, i.e. a model
in which the price of the shares is given as an input. We deduce the
time evolution of the portfolio of the various traders of the
market, as well as of other {\em observable} quantities. As in a
previous paper, we solve the equations of motion by means of a {\em
fixed point like} approximation.

\end{abstract}

\vspace{1cm}


\vfill

\newpage

\section{Introduction and motivations}

In some recent papers, \cite{bag1,bag2}, we have discussed why and
how a quantum mechanical framework, and in particular operator algebras and the number representation, can be used in the analysis of
 some simplified models of stock markets. In these models, among the other simplifications, no financial derivative are considered at all. For this reason our interest is different from that  discussed in \cite{baa}, even if the frameworks appear to be very close.
The main reason for using operator algebras in the analysis of this simplified
closed stock market comes from the following considerations: in the
closed market we have in mind the total amount of cash stays constant. Also, the
total number of shares does not change with time. Moreover, when a
trader $\tau$ interacts with a second trader $\sigma$, they change
money and shares in a {\em discrete fashion}: for instance,  $\tau$ increments his shares of 1 unit while his cash
decrements of a certain number of {\em monetary units} (which is the
minimum amount of cash existing in the market: 1 cent of dollar, for
example), which is exactly the price of the share. Of course, for the trader $\sigma$ the situation is just
reversed. So we have at least two quantities, the cash and the
number of shares, which change as multiples of two fixed
quantities. We further have two other quantities in our simplified
market: the price of the share (in our simplified market the traders
can exchange just a single kind of shares!) and the {\em market
supply}, i.e. the overall tendency of the market to sell a share.
For the same reason just discussed it is clear that also the price
of the share must change discontinuously. It is therefore convenient
to assume that also the market supply is labeled by a discrete quantity. In our naive
version of a stock market the price of the share is  related
to the market supply in a very direct way: if the market supply increases, then the value
of the share must decrease and viceversa. These are all the ingredients of our toy model of a stock market.

Operator algebras and quantum mechanics produce a very natural
settings for discussing such a system. Indeed they produce a natural
way for: (a) describing quantities which change with discrete steps;
(b) obtaining the differential equations for the relevant variables
of the system under consideration, the so-called {\em observables of
the system}; (c) finding conserved quantities. It is well known indeed that  operators satisfying the canonical commutation relations can be used quite naturally to discuss point (a) above. Hence the use of the Heisemberg dynamics, which is somehow intrinsic in quantum mechanics, appears a natural (but not exclusive!) choice to discuss points (b) and (c).

For these reasons we have suggested in \cite{bag1,bag2}  an operator-valued scheme for the description of such a simplified market. We refer to the Appendix for some preliminary result.

It should be mentioned that our idea is not entirely new. Indeed, the use of quantum mechanical tools, and more in general of {\em standard} statistical techniques  in the analysis of stock markets was already proposed by many authors, see for instance \cite{baa,mar,scha1,sasa} and
references therein. However, the approach discussed here and in \cite{bag1,bag2} is the first one, in our knowledge, in which the mechanism of the {\em exchange between traders} is given in terms of a boson-like based hamiltonian operator. This is close but somehow different from what is discussed in \cite{baa}, where a quantum hamiltonian $H_B$ is introduced in the analysis of financial derivatives which is undertaken by looking at the Schr\"odinger equation arising from $H_B$.

It should also be mentioned that we are still very far from being able to deal with a real stock market, and this is the reason why our analysis still misses of a detailed comparison with existing methods, which will be undertaken on some more realistic model.

\vspace{2mm}

The paper is organized as follows:

In  Section II we introduce the model originally proposed in
\cite{bag2}, and we deduce the related equations of motion.

In Section III we modify this model by assuming different
behaviors for the price of the share considered into the differential equations
as an independent input. Then we  discuss how the value of
the portfolio of a given trader changes with time, analyzing some numerical results while our
conclusions are given in Section IV. To keep the paper self-contained we have included in Appendix few useful and standard results on the number representation.

\section{The {\em all-in-one} model}

In \cite{bag2} we have introduced the following hamiltonian to
describe the time evolution of some {\em observables} of a toy model of
stock market consisting in $L$ traders exchanging shares of a single type, whose price is decided by the market itself, see below,

 \be
\left\{
\begin{array}{ll}
H=H_0+ \lambda\,H_I, \mbox{ where }  \\
H_0 = \omega_a\, \hat n+\omega_c\, \hat k+\omega_p \hat
P+\sum_{k\in\Lambda}\left(\Omega_A(k)\,\hat N_k+
 \Omega_C(k)\,\hat K_k+\Omega_O(k)\,\hat  O_k\right)\\
 H_I = \left( z^\dagger \,Z(f)+z\,Z^\dagger(\overline f)\right)+(p^\dagger \,o(g)+p \,o^\dagger(\overline g)) \\
\end{array}
\right. \label{21} \en Here $\omega_a$, $\omega_c$ and $\omega_p$
are positive real numbers and $\Omega_A(k)$, $\Omega_C(k)$ and
$\Omega_O(k)$ are real valued positive functions. They define
the {\em free} time evolution of the different operators of the market and $\lambda$ is a parameter measuring the strenght of the interactions between the different traders.
Moreover, introducing the operators $a$, $c$, $p$, $o_j$, $A_j$, $C_j$, $j\in\Lambda$, and their adjoints, which are assumed to satisfy the following commutation rules
\be [c,c^\dagger]=[p,p^\dagger]=[a,a^\dagger]=\1,\qquad
[o_i,o_j^\dagger]=[A_i,A_j^\dagger]=[C_i,C_j^\dagger]=\delta_{i,j}\1,\label{23}\en
the operators appearing in (\ref{21}) are defined as $\hat n=a^\dagger a$, $\hat k=c^\dagger
c$, $\hat P=p^\dagger p$, $\hat N_k=A^\dagger_k A_k$, $\hat
K_k=C^\dagger_k C_k$ and $\hat O_k=o^\dagger_k o_k$. In $H$ we have also
introduced the following {\em smeared fields}, depending on two
regular functions $f(k)$ and $g(k)$,
\be\left\{\begin{array}{ll}Z(f)=\sum_{k\in\Lambda}Z_k\,f(k)=\sum_{k\in\Lambda}A_k\,{C_k^\dagger}^{\,\hat
P}\,f(k), \\ Z^\dagger(\overline
f)=\sum_{k\in\Lambda}Z_k^\dagger\,\overline{f(k)}=\sum_{k\in\Lambda}A_k^\dagger\,{C_k}^{\,\hat
P}\,\overline{f(k)}\\
o(g)=\sum_{k\in\Lambda}o_k\,g(k) \\ o^\dagger(\overline
g)=\sum_{k\in\Lambda}o_k^\dagger\,\overline{g(k)},\end{array}\right.\label{22}\en
as well as the operators $z=a\,{c^\dagger}^{\,\hat P}$,
$Z_k=A_k\,{C_k^\dagger}^{\,\hat P}$ and their conjugates. Notice that we have separated here our stock market into two main components: the single trader $\tau$, which is arbitrary but fixed, and all the other traders, $\{\sigma_k\}$, whose set we call $\R$.
We refer to Table 1 below for a list of these operators and of their {\em economical meaning}, which can be deduced by the discussion following equation (\ref{a4}).

\vspace{5mm}
{\hspace{-.8cm}
\begin{tabular}{|c||c||c||c||c|} \hline    &the operator and.. &...its economical meaning               \\
\hline  & $a$  & annihilates a share in the portfolio of $\tau$    \\
\hline   & $a^\dagger$ & creates a share in the portfolio of $\tau$   \\
\hline  & $\hat n=a^\dagger a$ &counts the number of shares in the portfolio of $\tau$   \\
\hline\hline
\hline  & $c$  & annihilates a monetary unit in the portfolio of $\tau$    \\
\hline   & $c^\dagger$ & creates a monetary unit in the portfolio of $\tau$   \\
\hline  & $\hat k=c^\dagger c$ &counts the number of monetary units in the portfolio of $\tau$   \\
\hline\hline
\hline  & $p$  & it lowers the price of the share of one unit of cash   \\
\hline   & $p^\dagger$ & it increases the price of the share of one unit of cash   \\
\hline  & $\hat P=p^\dagger p$ &gives  the  value of the share   \\
\hline\hline
\hline  & $A_k$  & annihilates a share in the portfolio of $\sigma_k$    \\
\hline   & $A_k^\dagger$ & creates a share in the portfolio of $\sigma_k$   \\
\hline  & $\hat N_k=A_k^\dagger A_k$ &counts the number of shares in the portfolio of $\sigma_k$   \\
\hline\hline
\hline  & $C_k$  & annihilates a monetary unit in the portfolio of $\sigma_k$    \\
\hline   & $C_k^\dagger$ & creates a monetary unit in the portfolio of $\sigma_k$   \\
\hline  & $\hat K_k=C_k^\dagger C_k$ &counts the number of monetary units in the portfolio of $\sigma_k$   \\
\hline\hline
\hline  & $o_k$  & it lowers the k-th component of the market supply of one unit    \\
\hline   & $o_k^\dagger$ & it increases the k-th component of the market supply of one unit  \\
\hline  & $\hat O_k=o_k^\dagger o_k$ &gives  the  value of the the k-th component of the market supply   \\
\hline\hline

\end{tabular}

\vspace{4mm}

\indent
Table 1.-- List of operators and of their {\em economical} meaning.}

\vspace{5mm}

Notice now that, because of (\ref{23}), the following commutators can also be deduced, \cite{bag2}:
\be [\hat
K_k,C_q^{\,\hat P}]=-\hat P\,C_q^{\,\hat P}\,\delta_{k,q},\quad
[\hat K_k,{C_q^\dagger}^{\,\hat P}]=\hat P\,{C_q^{\dagger}}^{\,\hat
P}\,\delta_{k,q} \label{24}\en
To Table 1 we add now for completeness the economical meaning of, e.g., the operator $z$. This is rather evident, indeed.
Let us now consider a fixed  number vector, extending the one in (\ref{a4}),
\be
\varphi_{n,k,M,\{N\};\{K\};\{O\}}:=\frac{{a^\dagger}^n\,{c^\dagger}^k\,{p^\dagger}^M\,{A_1^\dagger}^{N_1}\cdots
{A_L^\dagger}^{N_L}{C_1^\dagger}^{K_1}\cdots
{C_L^\dagger}^{K_L}{o_1^\dagger}^{O_1}\cdots{o_L^\dagger}^{O_L}}{\sqrt{n!k!M!N_1!\ldots
N_L!\,K_1!\ldots K_L!O_1!\ldots O_L!}}\,\varphi_0, \label{24bis}\en
where $\{N\}=N_1,N_2,\ldots,N_L$, $\{K\}=K_1,K_2,\ldots,K_L$,  $\{O\}=O_1,O_2,\ldots,O_L$ and  $\varphi_0$ is the vacuum of all the annihilation operators involved here.

The action of $z$ on such a vector destroys a share in the portfolio of the trader $\tau$ because of the presence of the annihilation operator $a$ and, at
the same time, creates as many monetary units as $\hat P$
prescribes because of ${c^\dagger}^{\,\hat P}$. Of course, in the interaction hamiltonian $H_I$ such an operator is associated to
$Z^\dagger(\overline f)$ which acts exactly in the opposite way on
the other traders $\{\sigma_k\}$ of the market: one share is created in the cumulative portfolio of $\R$ while
$\hat P$ {\em quanta} of money are destroyed, since they are used to
pay for the share. This interpretation clearly follows from the
commutation rules in (\ref{23}) and (\ref{24}). That's why $z$ is called a {\em selling} and $Z^\dagger(\overline f)$  a {\em buying operator}.

Finally, if $L+1$ is the total number of traders of our market, then
the cardinality of $\Lambda$, which is a subset of $\Bbb{N}$ which
labels the traders of $\R$, is obviously $L$.

The main aim of our analysis is to recover the equations of motion
for the portfolio of the trader $\tau$ defined as \be \hat
\Pi(t)=\hat P(t)\,\hat n(t)+\hat k(t).\label{26}\en This is a natural definition, since it is just the sum of the cash and of the total value of the shares that $\tau$ possesses at time $t$. Once again, we stress that in our simplified model no room is given to the financial derivatives, and not even to any different and realistic mechanism but a simple exchange cash$\leftrightarrow$shares. This oversimplification makes  our model reasonably simple to be analyzed using standard perturbative techniques but, at the same time, strongly limits the validity of the model itself. However, to our knowledge, this is a  feature shared by most of the models of stock markets existing in the literature, which are usually focused just on some particular aspect of the market and not, because of its complexity, to the whole system.

\vspace{2mm} {\bf Remarks:--} (1) Table 1 has a physical
origin which is discussed in \cite{bag2} where the so-called
stochastic limit approach for quantum open systems was adopted. The role of the smeared field in this context has been clarified. It may be worth stressing that what we call stochastic limit approach is essentially a  perturbative technique originally discussed in a quantum mechanical context, \cite{acc}.

(2) Of course, since $\tau$ can be chosen arbitrarily, the asymmetry
of the model is just apparent. In fact, changing $\tau$, we will be
able, in principle, to find the time evolution of the portfolio of
each trader of the stock market.

\vspace{2mm}

The interpretation suggested above concerning $z$ and $Z(f)$ is
also based on the following result: let us define the operators \be
\hat N:=\hat n+\sum_{k\in\Lambda}\,\hat N_k,\quad \hat K:=\hat
k+\sum_{k\in\Lambda}\,\hat K_k,\quad \hat \Gamma:=\hat
P+\sum_{k\in\Lambda}\hat O_k\label{25}\en  We have seen in
\cite{bag2} that  $\hat N$, $\hat K$ and $\hat \Gamma$ are constants
of motion: $[H,\hat N]=[H,\hat K]=[H,\hat \Gamma]=0$. This is in agreement with our interpretation: we are
constructing a closed market in which the total amount of money and
the total number of shares are preserved and in which, if the total
supply increases, then the price of the share must decrease in such a way that
 $\hat \Gamma$ stays constant.

In \cite{bag2}
we have shown how to use the stochastic limit approach or a fixed
point-like (fpl) procedure to get suitable approximations of $\hat\Pi(t)$ in (\ref{26}).
Here we just consider this last approach, since it is more relevant
for the analysis we will perform here. We refer to
\cite{bag2} for the details of our derivation, and for the details
on the assumptions that have been taken along the way. Here we just
write down the system of operatorial differential equations which we
have deduced and which looks like

\be\left\{\begin{array}{lll}\frac{d\hat
n(t)}{dt}=i\lambda\left(-z^\dagger(t)\,Z(f,t)+z(t)\,Z^\dagger(\overline
f,t)\right),\\
\frac{d\hat k(t)}{dt}=i\lambda\,
P_c(t)\,\left(z^\dagger(t)\,Z(f,t)-z(t)\,Z^\dagger(\overline
f,t)\right),\\
\frac{dz(t)}{dt}=i\left( P_c(t)\omega_c-\omega_a\right)\,z(t)+i\lambda[z^\dagger(t),z(t)]\,Z(f,t),\\
\frac{dZ(f,t)}{dt}=i\,Z\left((
P_c(t)\Omega_C-\Omega_A)f,t\right)+i\lambda\,z(t)\,[Z^\dagger(\overline
f,t),
Z(f,t)],\\
\end{array}\right.\label{27}\en
where \be P_c(t)=\frac{1}{2}\left[(M+O)+(M-O)\cos(2\lambda t)\right]
\label{28}\en is the mean value of the operator $\hat P(t)$ on a
vector state defined by (\ref{24bis}), and which we
simply indicate with $\omega$, and can be explicitly found due to
the simple expression of $H$, \cite{bag2}: $P_c(t)=\omega(\hat P(t))$. Here $M$ and $O$ are respectively the
initial price of the shares and the initial value of the market
supply (which is related to the set $\{O\}$ in (\ref{24bis})). These equations produce the dynamical behavior of our market
 after taking their mean value on the number vector state
$\omega$, \cite{bag2}. As  mentioned in Appendix, this state
has a double effect: first of all, it allows us to move from the
{\em quantum} dynamics of the model to its {\em classical}
counterpart, since we use $\omega$ to replace the time dependent
operators with their mean values, which are ordinary functions of
time. Secondly, such a vector state is used to fix the initial conditions
of the differential equations, that is the initial number of shares,
the initial cash and so on.

In order to simplify  the analysis of this system we have assumed in
\cite{bag2} that both $\Omega_C(k)$ and $\Omega_A(k)$ are constant
for $k\in\Lambda$. Indeed, under this assumption, the last two
equations in (\ref{27}) form by themselves a closed system of
differential equations in the non abelian variables $z(t)$ and
$Z(f,t)$: \be\left\{\begin{array}{lll}
\frac{dz(t)}{dt}=i\left( P_c(t)\omega_c-\omega_a\right)\,z(t)+i\lambda\,Z(f,t)\,[z^\dagger(t),z(t)],\\
\frac{dZ(f,t)}{dt}=i\,(P_c(t)\Omega_C-\Omega_A)\,Z(f,t)+i\lambda\,z(t)\,[Z^\dagger(\overline
f,t),Z(f,t)].\\
\end{array}\right.\label{210}\en

Getting the exact solution of the system (\ref{27}), with
(\ref{210}) replacing the two last equations, is a hard job. However, a
 natural approximation can be constructed considering the fpl
 approach for which, again, we refer to \cite{bag2}. It should be mentioned, however, that for the time being we have not undertaken the rigorous analysis of such an approximation (i.e. the estimate of the error), which is adopted here mainly because it can be very easily implemented and, more than this, because it is quite natural.  The result is
 the following: $z(t)$ and $Z(f,t)$ can be approximated by two
 different (and not commuting) operators $z_1(t)$ and $Z_1(f,t)$, \cite{bag2}:
\be z_1(t)=z\,\eta_1(t)+Z(f)\,[z^\dagger,z]\,\eta_2(t),\quad
Z_1(f,t)=Z(f)\,\tilde\eta_1(t)+z\,[Z(\overline
f)^\dagger,Z(f)]\,\tilde\eta_2(t), \label{211}\en where
\be\left\{\begin{array}{ll} \eta_1(t)=1+i\int_0^t(
P_c(t')\omega_c-\omega_a)\,e^{i\chi(t')}\,dt',\quad
\eta_2(t)=i\lambda\int_0^te^{i\tilde\chi(t')}\,dt'\\
\tilde\eta_1(t)=1+i\int_0^t(
P_c(t')\Omega_C-\Omega_A)\,e^{i\tilde\chi(t')}\,dt',\quad
\tilde\eta_2(t)=i\lambda\int_0^te^{i\chi(t')}\,dt'
\end{array}\right.\label{212}
\en and $\chi(t)=\alpha t+\beta \sin(2\lambda t)$,
$\tilde\chi(t)=\tilde\alpha t+\tilde\beta \sin(2\lambda t)$, where
$\alpha=\frac{1}{2}((M+O)\omega_c-2\omega_a)$,
$\tilde\alpha=\frac{1}{2}((M+O)\Omega_C-2\Omega_A)$,
$\beta=\frac{\omega_c}{4\lambda}(M-O)$,
$\tilde\beta=\frac{\Omega_C}{4\lambda}(M-O)$.

It is now easy to find that the mean values of the first two
equations in (\ref{27}) on the state $\omega$ can be written as
\be\left\{\begin{array}{lll}\dot n(t)=\frac{d
n(t)}{dt}=-2\lambda\Im\left\{\omega\left(z(t)\,Z^\dagger(\overline
f,t)\right)\right\},\\
\dot k(t)=\frac{d
k(t)}{dt}=2\lambda\,P_c(t)\Im\left\{\omega\left(z(t)\,Z^\dagger(\overline
f,t)\right)\right\},\\
\end{array}\right.\label{213}\en
which in particular implies that $P_c(t)\dot n(t)+\dot k(t)=0$ for
all $t$. Hence we find that $\dot\Pi(t)=\dot P_c(t)\,n(t)$.
It should be remarked that, because of this relation, since $M=O$ in
(\ref{28}) implies $P_c(t)=P_c(0)=M$, then when $M=O$ the dynamics
of the portfolio of $\tau$ is trivial, $\Pi(t)=\Pi(0)$, even if both
$n(t)$ and $k(t)$ may change in time.

Let us now insert $z_1(t)$ and $Z_1(f,t)$ in equations (\ref{213}).
If $\omega$ is the usual number state, and if we call
\be\left\{\begin{array}{lll}\omega(1):=\omega\left(zz^\dagger\,[Z^\dagger(\overline
f),Z(f)]\right\},\\
\omega(2):=\omega\left(Z(f)Z^\dagger(\overline
f)\,[z^\dagger,z]\right\},\\
r(t)=\omega(1)\,\eta_1(t)\,\overline{\tilde\eta_2(t)}+\omega(2)\,\eta_2(t)\,\overline{\tilde\eta_1(t)}\\
\end{array}\right.\label{214}\en
then we get \be\left\{\begin{array}{lll} n(t)=n-2\,\lambda\Im\left\{\int_0^t r(t')\,dt'\right\},\\
k(t)=k+2\,\lambda\Im\left\{\int_0^t P_c(t')\,r(t')\,dt'\right\}.\\
\end{array}\right.\label{214b}\en

The time dependence of the portfolio can now be written as \be
\Pi(t)=\Pi(0)+\delta\Pi(t),\label{214c}\en with $$
\delta\Pi(t)=n(O-M)\sin^2(\lambda t)+$$
\be+\left(-2\lambda\,\Im\left\{\int_0^t
r(t')\,dt'\right\}\left(M+(O-M)\sin^2(\lambda
t)\right)+2\lambda\,\Im\left\{\int_0^t
P_c(t')\,r(t')\,dt'\right\}\right), \label{215}\en which gives the
variation of the portfolio of $\tau$ in time. \vspace{2mm}

\noindent{\bf Remark:} it is worth noticing that $\delta\Pi(t)=0$
for all $t$ if $\lambda=0$. This is reasonable, since $\lambda=0$
means no interaction in (\ref{21}) and, as a consequence, no way to
change the original status quo.

\vspace{2mm}

In \cite{bag2} we have discussed some particular solutions of this
system under special conditions. In particular we have shown that
there exist situations which are rather {\em convenient} for the trader
$\tau$, meaning with this that  $\delta\Pi(t)\geq0$ for
all $t\geq0$: under these conditions, within our framework and our approximations, the
total value of the portfolio of $\tau$ {\bf never} decreases! These
conditions are, not surprisingly, related to the constants which
enter in the definition of $H$. However, due to the simplified
mechanism which fixes the price of the share, it is clear that we
cannot trust very much these conclusions: as already stated, the model still needs
further improvements. In particular different kind of shares and a
more realistic dynamics for the price must still be implemented.
Both these extensions require some care, and will be considered in a series of
papers which are now in preparation. We consider a first generalization in the next section, which is used to get a
deeper insight on the role of the constants appearing in the
definition of the model itself.

\section{The mixed model}

In this section we  discuss in some details a different model of a
stock market, based again on almost all the same  assumptions
introduced in \cite{bag1,bag2}. We call this  a {\em mixed model}
since it has an {\em hamiltonian ingredient}, which involves again the
cash and the number of shares of both $\tau$ and of the traders in
$\R$, and an {\em empirical part}, since the dynamics of the price
is not deduced as before from the market
supply but simply given as an input in  (\ref{27}), which  is here replaced by the following system
\be\left\{\begin{array}{lll}\frac{d\hat
n(t)}{dt}=i\lambda\left(-z^\dagger(t)\,Z(f,t)+z(t)\,Z^\dagger(\overline
f,t)\right),\\
\frac{d\hat k(t)}{dt}=i\lambda\,
P(t)\,\left(z^\dagger(t)\,Z(f,t)-z(t)\,Z^\dagger(\overline
f,t)\right),\\
\frac{dz(t)}{dt}=i\left( P(t)\omega_c-\omega_a\right)\,z(t)+i\lambda\,Z(f,t)\,[z^\dagger(t),z(t)],\\
\frac{dZ(f,t)}{dt}=i\,(P(t)\Omega_C-\Omega_A)\,Z(f,t)+i\lambda\,z(t)\,[Z^\dagger(\overline
f,t),Z(f,t)].\\
\end{array}\right.\label{31}\en
Hence here $P(t)$ is a classical {\em external field}, for which we will consider different analytical expressions in the rest of the paper.

The main reason why we are considering  this simplified model is that the models proposed in \cite{bag1,bag2} and in the
previous section, all depend on a set of parameters whose role in the description of our toy stock market needs
to be yet fully understood. At a preliminary stage it is therefore convenient to reduce the number of {\em degrees of freedom} and to deal with the simpler model described by (\ref{31}).
This should produce a deeper understanding which will be useful to construct other and more
realistic models. This analysis is also necessary if we want to compare our analysis with the ones already existing in the literature, where the role of certain parameters is quite often absolutely crucial, see \cite{mlux,alfi} for instance. The relation between these parameters, for instance, decides the nature of the traders which form our market, \cite{alfi}, or how these traders react to the dynamics of the market.

Needless to say, a realistic model of a stock
market should contain, first of all, more than a single kind of
share. Moreover, a mechanism which fixes the prices of all these
shares is needed, and this is surely not a trivial task since
interactions between the various traders and the various kind of
shares must be taken into account, as well as external sources of information. This is exactly our final aim:
produce a model which contains $N$ traders and $L$ different kind of
shares whose prices are decided by some {\em global} mechanism to be
identified. However, for the time being, we just continue
our preliminary analysis, paying particular attention to the role of
the parameters of the simplified model described
here.

\vspace{2mm}

{\bf Remark:--}
One could think to deduce system (\ref{31}) from the following
effective (time-dependent) hamiltonian $H_{eff}$ which looks very
similar to that in (\ref{21})
 \be
\left\{
\begin{array}{ll}
H_{eff}(t)=H_{eff,0}+ \lambda\,H_{eff,I}(t), \mbox{ where }  \\
H_{eff,0} = \omega_a\, \hat n+\omega_c\, \hat
k+\sum_{k\in\Lambda}\left(\Omega_A\,\hat N_k+
 \Omega_C\,\hat K_k\right)\\
 H_{eff}(t) = \left( z^\dagger(t) \,Z(f,t)+z(t)\,Z^\dagger(\overline f,t)\right)), \\
\end{array}
\right. \label{32} \en where, for instance,
$Z(f,t)=\sum_{k\in\Lambda}Z_k\,f(k)=\sum_{k\in\Lambda}\,f(k)\,A_k\,{C_k^\dagger}^{\,
P(t)}$. However this is not rigorous because the time dependence of
$P(t)$ modifies the standard Heisenberg equation
$\frac{d}{dt}X(t)=i[H,X(t)]$. Hence it is more convenient to take
(\ref{31}) as the starting point, and (\ref{32}) just as a formal
hamiltonian heuristically related to the economical system.

\vspace{2mm}

As for the numerical aspects, we have restricted here the analysis of the
model to a time interval $t\in[0,t_0]$, with $t_0=6$ just to fix the ideas, and we have considered the following forms for
$P(t)$: $P_1(t)=t$,
 $$P_2(t)=\left\{
\begin{array}{ll}
\!0, \mbox{ if } t\in[0,1]  \\
\!t-1, \mbox{ if } t\in[1,3],  \\
\!2, \mbox{ if } t>3 \\
\end{array}
\right.\,  P_3(t)=\left\{
\begin{array}{ll}
\!2, \mbox{ if } t\in[0,1]  \\
\!3-t, \mbox{ if } t\in[1,3],  \\
\!0, \mbox{ if } t>3  \\
\end{array}
\right.\,  P_4(t)=\left\{
\begin{array}{ll}
\!0, \mbox{ if } t\in[0,1]  \\
\!t-1, \mbox{ if } t\in[1,3]  \\
\!5-t, \mbox{ if } t\in[3,4]  \\
\!1, \mbox{ if } t>4.
\end{array}
\right.$$ Hence, we are considering four different situations:
$P_1(t)$ and $P_2(t)$ describe a not decreasing price, while
$P_3(t)$ is a not increasing function. Finally, $P_4(t)$  describes a share whose value increases up to a maximum and then
decreases and reaches a limiting value. The reason for using these
rather than other and more regular functions is that with these
choices it is easier to compute analytically some of the quantities
appearing in the solution of system (\ref{31}).

\vspace{2mm}

Let us now show quickly how to find this solution by means of the fpl approximation. For more details we refer
to \cite{bag2}. As in Section II the main idea is finding first an
approximated solution of the last two equations of (\ref{31}) and
then using these solutions in the first two equations of the same
system.  In this way
we recover the solution in (\ref{211})-(\ref{212}). In particular this last equation can be rewritten as
\be\left\{\begin{array}{ll} \eta_1(t)=e^{i\chi(t)},\quad
\eta_2(t)=i\lambda\int_0^te^{i\tilde\chi(t')}\,dt'\\
\tilde\eta_1(t)=e^{i\tilde\chi(t)},\quad
\tilde\eta_2(t)=i\lambda\int_0^te^{i\chi(t')}\,dt',
\end{array}\right.\label{33}
\en with $\chi(t)=\int_0^t(P(t')\omega_c-\omega_a)\,dt'$ and
$\tilde\chi(t)=\int_0^t(P(t')\Omega_C-\Omega_A)\,dt'$. These results
can be used in the first two equations in (\ref{31}) and we get \be
\left\{\begin{array}{ll} n(t)=n+\delta n(t)=n-2\lambda\int_0^t\Im(r(t'))\,dt'\\
k(t)=k+\delta k(t)=k+2\lambda\int_0^tP(t')\,\Im(r(t'))\,dt',\\
\end{array}\right.\label{34}
\en with obvious notation, having introduced the function $r(t)$ as
in (\ref{214}). The time evolution of the portfolio of $\tau$,
$\Pi(t)=\Pi(0)+\delta\Pi(t)$, can be written as \be
\delta\Pi(t)=n(P(t)-P(0))+ P(t)\,\delta n(t)+\delta
k(t)\label{35}\en which reduces to (\ref{215}) under the assumptions
of Section II.

In the rest of the paper we consider some simplifying  conditions, useful to display easily our results:
first of all, as in \cite{bag2}, we reduce the reservoir of the
market just to another trader, $\sigma$. This makes the computation
of $\omega(1)$ and $\omega(2)$ in (\ref{214}) rather easy and we get
\be\left\{
\begin{array}{ll}
\omega(2)=-|f(1)|^2(1+n')\frac{k!}{(k-M)!}\,\frac{k'!}{(k'-M)!}  \\
\omega(1)=\omega(2)+|f(1)|^2n'\frac{k!}{(k-M)!}\,\frac{(k'+M)!}{k'!},   \\
\end{array}
\right.\label{41}\en
which depends, as expected, on the initial values of the market.
As a second assumption, we just consider the case of a trader $\tau$
which is {\em entering into the market}, so that it possess no share
at all at $t=0$. Therefore we take $n=0$. This has an immediate
consequence: in order to have an economical meaning, $\delta n(t)$
can only be non negative! This is actually what happens in all the
examples we have considered so far.

\subsection{Numerical results and conclusions}

We discuss now the numerical results of
the solutions of system (\ref{31}) for the different choices of
$P(t)$ introduced previously. Here, for concreteness' sake, we fix
$f(1)=10^{-3}$ in (\ref{41}) and, as already mentioned, $n=0$, and we consider the following cases (different choices
of constants in $H$) and subcases (different initial conditions):

\vspace{2mm}

{\bf case I:} $(\omega_a,\omega_c,\Omega_A,\Omega_C) = (1,1,1,1)$;
{\bf case II:} $(\omega_a,\omega_c,\Omega_A,\Omega_C) = (10,10,1,1)$;
{\bf case III:}
$(\omega_a,\omega_c,\Omega_A,\Omega_C) = (1,1,10,10)$;
{\bf case IV:}
$(\omega_a,\omega_c,\Omega_A,\Omega_C) = (20,10,5,1)$;
{\bf case V:}
$(\omega_a,\omega_c,\Omega_A,\Omega_C) = (1,5,10,20)$;
{\bf case VI:}
$(\omega_a,\omega_c,\Omega_A,\Omega_C) = (1,3,2,7)$.

{\bf Subcase a:} $(k,k',n',M)=(20,20,100,2)$;
{\bf subcase b:}
$(k,k',n',M)=(80,20,100,2)$;
{\bf subcase c:} $(k,k',n',M)=(20,80,100,2)$;
{\bf subcase d:} $(k,k',n',M)=(80,80,100,2)$.

\vspace{3mm}

The first interesting result of our numerical computations is that
there are several situations, like for instance IIIa, Va and VIa for
$P_1(t)$, or IIIa, IVa, Va and VIa for $P_2(t)$, for which the
variation of $\delta n$ is larger than zero but always (i.e. for
$t<6$) strictly smaller than 1. Analogously there are other
situations in which $\delta k(t)$, belongs to the interval $]-1,1[$
for $t<6$, like in IIIa, Va and VIa for $P_1(t)$. This can be
interpreted as follows: since there is not much cash going around
the market, it is quite unlikely that some transaction may take
place between the two traders $\tau$ and $\sigma$. Figure I show
$\delta n(t)$ and $\delta k(t)$ for $P_1(t)$ in case Va.

\begin{center}
\mbox{\includegraphics[height=3.2cm, width=4.5cm]
{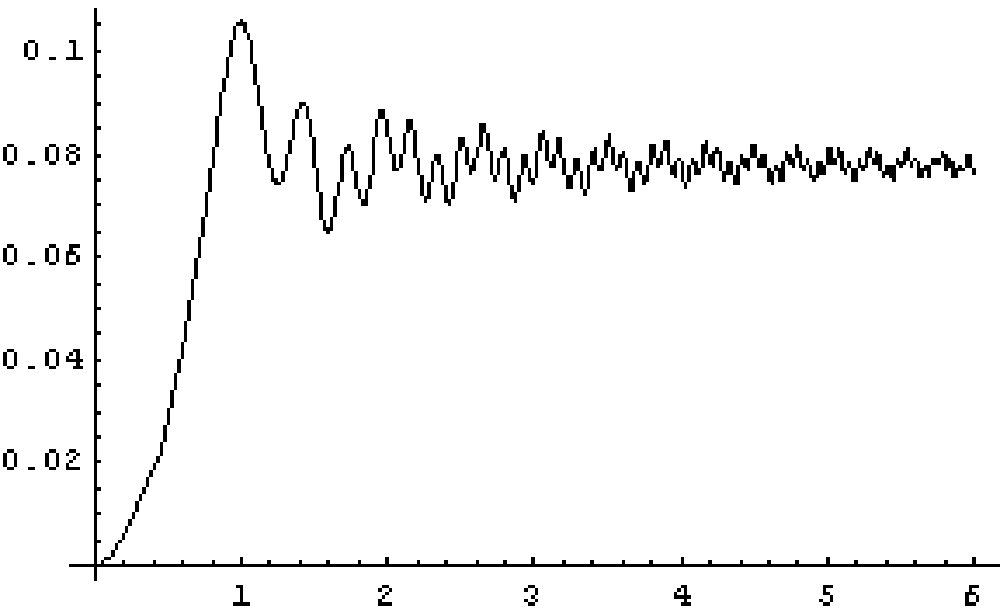}}\hspace{30mm}
\mbox{\includegraphics[height=3.2cm, width=4.5cm] {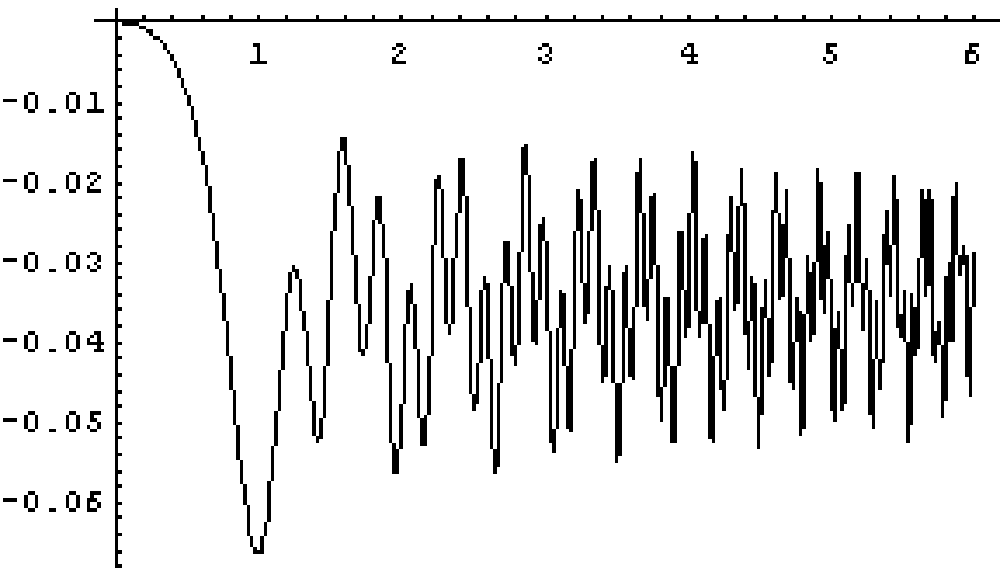}}\hfill\\
\begin{figure}[h]
\caption{\label{fig1} $\delta n(t)$ (left) and $\delta k(t)$ (right)
for $P_1(t)$, case Va: no transaction is possible for $t\in[0,6]$ }
\end{figure}
\end{center}

Moreover, there are situations in which the total quantity of cash
or the number of shares exceed their initial values, like for
instance in cases Id or IId for $P_2(t)$, see Figure 2 below. This
is clearly incompatible with the fact that the total amount of cash
and the total number of shares should be constant in time, and is
uniquely due to the fpl approximation which we have adopted to solve
the original system of differential equations. In other words, in
these cases there exists a value of $t$, $t_f<6$, such that for
$t>t_f$ the fpl approximation does not work anymore and our results become
meaningless. This happens usually in each subcase d, because of the
high values assumed by $\omega(1)$ and $\omega(2)$ due to the initial
conditions.  This problem could be solved if we adopt a better
approximation to solve the differential equations in (\ref{31}). This is part of the work in
progress, but is far from being an easy task, due to the non-commutative nature of the system (\ref{31}), which is also non linear. A more promising approach seems the use of time-dependent perturbation theory rather than the fpl approximation considered here. This is also work in progress.

\begin{center}
\mbox{\includegraphics[height=3.2cm, width=4.5cm]
{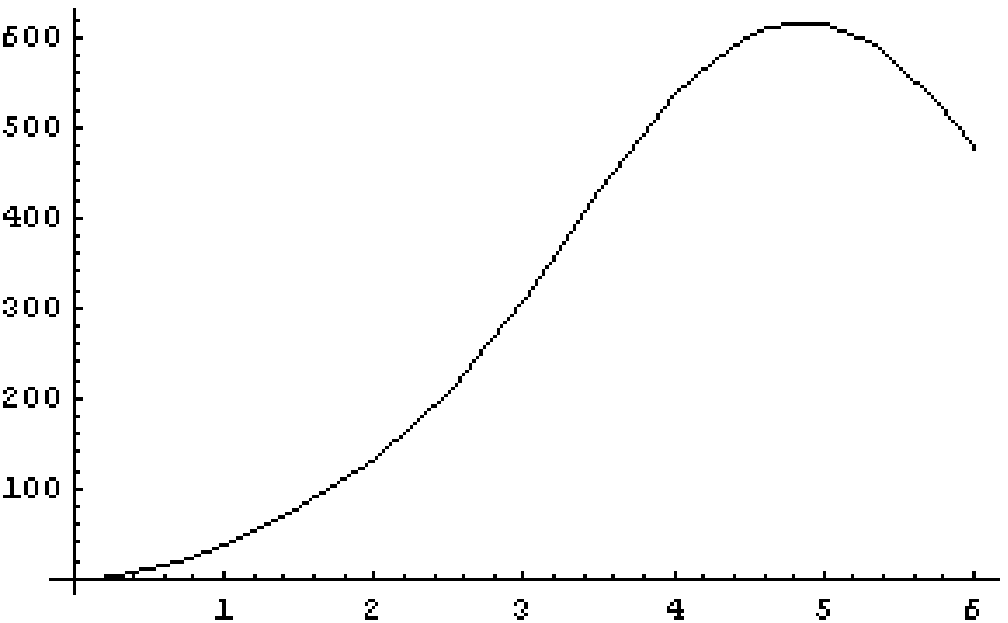}}\hspace{30mm}
\mbox{\includegraphics[height=3.2cm, width=4.5cm] {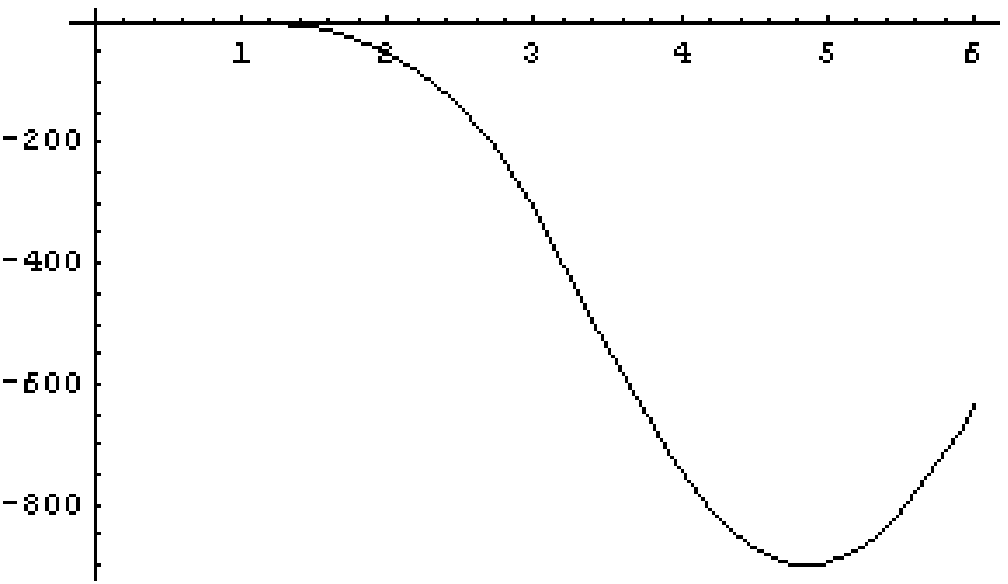}}\hfill\\
\begin{figure}[h]
\caption{\label{fig2} $\delta n(t)$, case Id (left) and $\delta k$,
case IId (right), for $P_2(t)$: the fpl approximation does not hold
in all of $[0,6]$ }
\end{figure}
\end{center}

As already mentioned, in all the $P(t)$ we have considered in this
paper we have found that, for all $t$ in our range, $\delta n(t)$ is
not negative, as it should since we are just considering the initial
condition $n=0$: hence the fpl approximation respects this feature
of the model. We also observe that the range of variations of all
the relevant observables, i.e. $\delta n(t)$, $\delta k(t)$ and
$\delta \Pi(t)$, is minimum in subcases a, maximum in the subcases d,
while is intermediate for both subcases b and c: this is clearly due to the
fact that different initial conditions imply different values of
$\omega(1)$ and $\omega(2)$, which for instance increase very fast with $k$ and
$k'$, see (\ref{41}). We further observe that the analytic
behavior of $\delta\Pi(t)$ is close (sometimes very close) to that
of $P(t)$. This is shown in  Figure 3, which is
obtained considering $P(t)=P_4(t)$: in the upper figure we plot
$P_4(t)$, while the three figures down shows $\delta\Pi(t)$ in case
Ic (left),  IVd (center) and  Vc (right). From these figures it
appears clearly that when the price of the share stays constant, there
is no variation of $\delta\Pi(t)$. On the contrary, this increases
when $P(t)$ increases and decreases when $P(t)$ decreases. This is
not a peculiar feature of this example but can be analytically
deduced from formula (\ref{35}) with $n=0$. Indeed, if we just
compute the time derivative of both side and use the definitions of
$\delta n(t)$ and $\delta k(t)$, we find that $\dot\Pi(t)=\dot
P(t)n(t)$ and, since $n(t)$ is non negative, $\dot\Pi(t)$ and $\dot
P(t)$ have the same sign. Figure 3 also shows that there are
conditions (Vc, for instance), in which $\delta\Pi(t)$ looks very
much the same as $P(t)$, while with different choices of the
constant of the hamiltonian and of the initial conditions the shape
of $\delta\Pi(t)$  appear sufficiently different from that of
$P(t)$. Of course this is an interesting feature of the model but is also a measure of how oversimplified the real system is here. A more realistic result could be found, maybe, introducing other mechanisms which are missing in our toy model. We will briefly come back to this point in the next section.

\vspace*{2mm}

\begin{center}
\mbox{\includegraphics[height=3.2cm, width=4.5cm]
{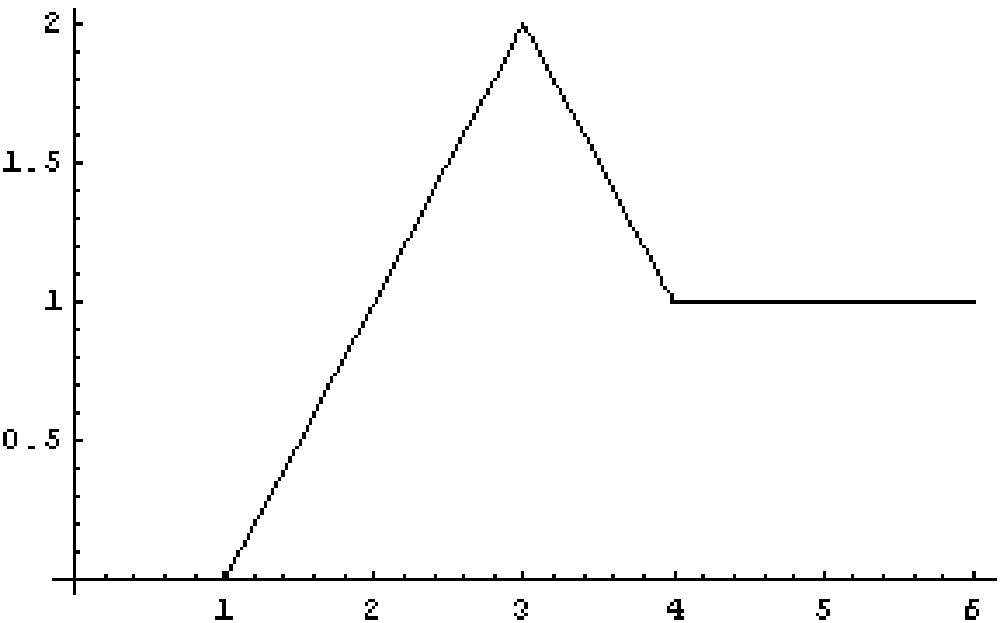}}\hfill\\
\mbox{\includegraphics[height=3.2cm, width=4.5cm] {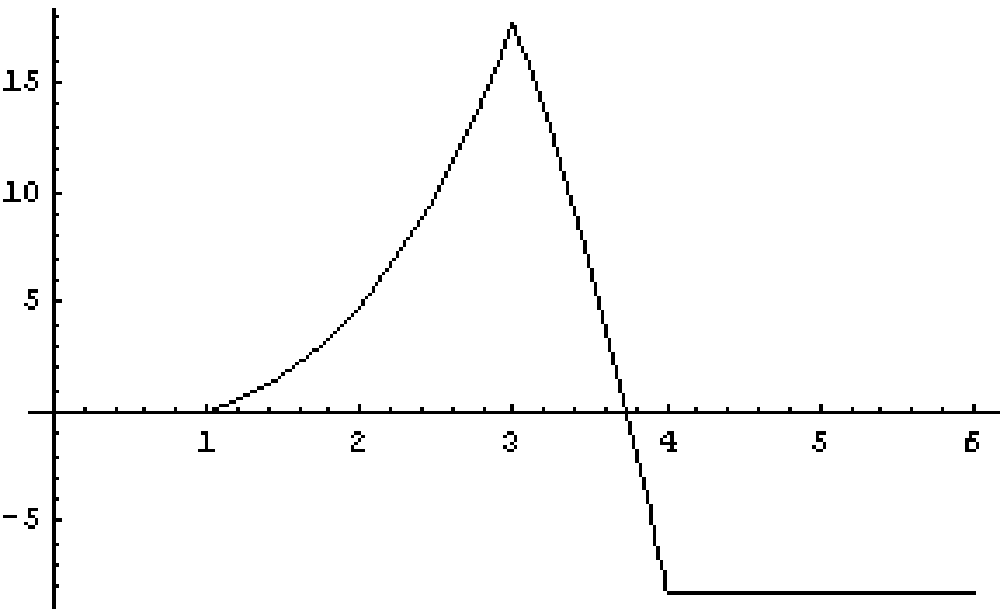}}
\hspace{6mm} \mbox{\includegraphics[height=3.2cm, width=4.5cm]
{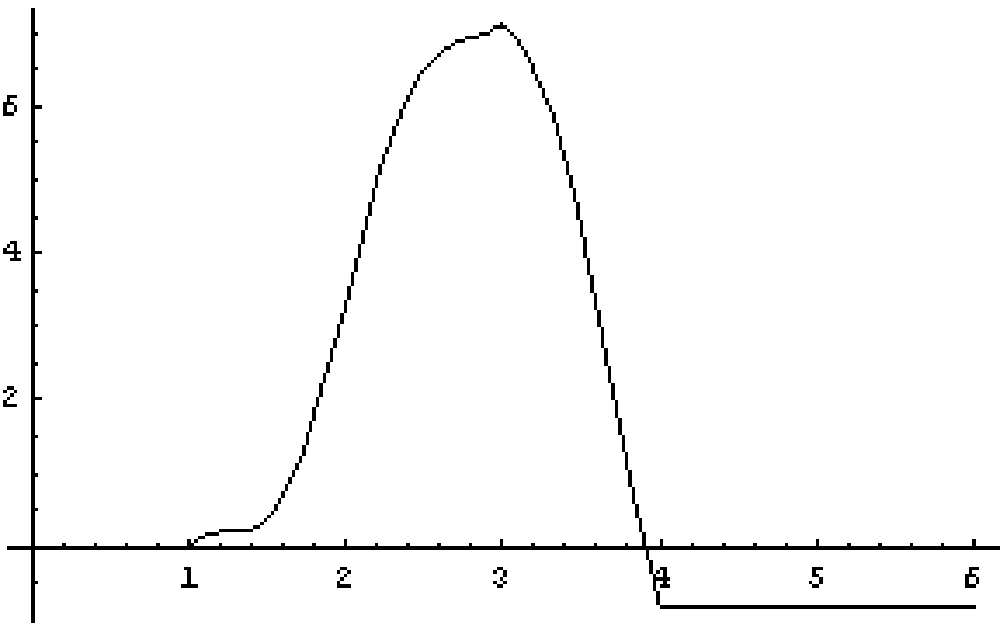}}\hspace{6mm}
\mbox{\includegraphics[height=3.2cm, width=4.5cm] {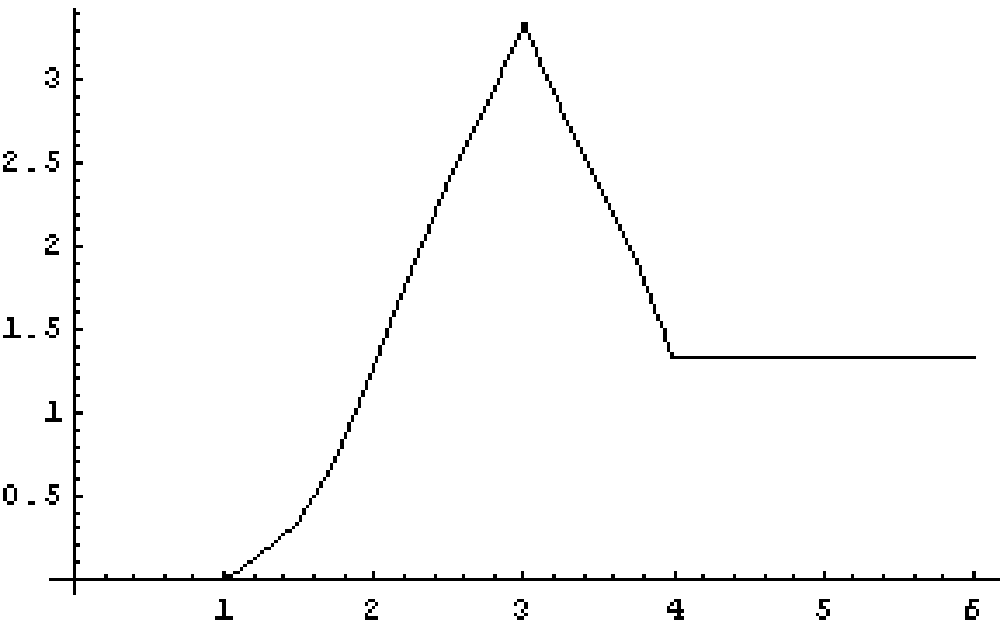}}\hfill\\
\begin{figure}[h]
\caption{\label{fig3}up: $P_4(t)$; down: $\delta\Pi(t)$ in cases Ic
(left), IVd (middle), Vc (right)  }
\end{figure}
\end{center}

We end this preliminary analysis by focusing our attention on the variation of
the portfolio of $\tau$: in the following table we show the range of
variation of $\delta\Pi(t)$ for the different cases, subcases and
choices of $P(t)$. Notice that the numbers are deduced from the
plots, so that they are only indicative.

\begin{tabular}{|c||c||c||c||c|} \hline    &$P_1(t)$ &$P_2(t)$  &$P_3(t)$  &$P_4(t)$            \\
\hline {\bf I} a & from 0 to 4  & from 0 to 1.2  & from -1.2 to 0 & from -0.5 to 1.25 \\
\hline  {\bf I} b & from 0 to 60 & from 0 to 17.5 & from -17.5 to 0  &from -7 to 17\\
\hline {\bf I} c & from 0 to 70 & from 0 to 20  & from -20 to 0 &from -8 to 20\\
\hline {\bf I} d & from 0 to 900 & from 0 to 300 & from -300 to 0 &from -110 to 300 \\
\hline\hline
\hline {\bf II} a & from 0 to 3.5  & from 0 to 1  & from -1 to 0 & from -0.4 to 1\\
\hline {\bf II} b & from 0 to 55 & from 0 to 17 & from -16 to 0  &from -7 to 17\\
\hline {\bf II} c & from 0 to 55 & from 0 to 17.5  & from -17.5 to 0 &from -7 to 17\\
\hline {\bf II} d & from 0 to 900 & from 0 to 270 & from -280 to 0 & from -100 to 250\\
\hline \hline
\hline {\bf III} a & from 0 to 1  & from 0 to 0.35  & from -0.3 to 0 & from -0.1 to 0.3\\
\hline {\bf III} b & from 0 to 8 & from 0 to 2.5 & from -2.5 to 0  &from -0.5 to 2.5\\
\hline {\bf III} c & from 0 to 17.5 & from 0 to 5.2  & from -5.2 to 0 &from -2 to 5.3\\
\hline {\bf III} d & from 0 to 140 & from 0 to 43 & from -43 to 0  &from -10 to 45\\
\hline \hline
\hline {\bf IV} a & from 0 to 0.7  & from 0 to 0.025  & from -0.15 to 0 & from -0.01 to 0.03\\
\hline {\bf IV} b & from 0 to 9 & from 0 to 0.4 & from -1.4 to 0  &from -0.05 to 0.4\\
\hline {\bf IV} c & from 0 to 10 & from 0 to 0.45  & from -2.5 to 0 &from -0.2 to 0.42\\
\hline {\bf IV} d & from 0 to 150 & from 0 to 7 & from -22 to 0  &from -0.4 to 7\\
\hline \hline
\hline {\bf V} a & from 0 to 0.4  & from 0 to 0.2  & from -0.03 to 0 &from 0 to 0.2 \\
\hline {\bf V} b & from 0 to 6  & from 0 to 1.25  & from -0.4 to 0  &from 0 to 1.3\\
\hline {\bf V} c & from 0 to 7  & from 0 to 3.2  & from -0.5 to 0 &from 0 to 3.5\\
\hline {\bf V} d & from 0 to 100  & from 0 to 22  & from -7 to 0  &from 0 to 22\\
\hline \hline
\hline {\bf VI} a & from 0 to 0.6  & from 0 to 0.7  & from -0.06 to 0 &from 0 to 0.7 \\
\hline {\bf VI} b & from 0 to 8  & from 0 to 9  & from -0.6 to 0  &from 0 to 9\\
\hline {\bf VI} c & from 0 to 10  & from 0 to 12  & from -1 to 0 &from 0 to 12\\
\hline {\bf VI} d & from 0 to 120  & from 0 to 150  & from -11 to 0  &from 0 to 150\\
\hline

\end{tabular}

\vspace{4mm}

\indent
Table 2.-- Range of
variation of $\delta\Pi(t)$ for different choices of $P_j(t)$.

\vspace*{5mm}

We see that for all given subcase, case I is the one in which
we find the widest range of variation of $\delta\Pi(t)$. In particular, for
$P_j(t)$ with $j=1,2,4$, $\delta\Pi(t)$ increases its value more
than in the other cases (II-VI) for any fixed subcase. If $j=3$ then
$\delta\Pi(t)$ decreases its value in case I more than in the other
cases (II-VI). This difference between $P_3(t)$ and the others $P_j(t)$, $j\neq 3$, is clearly related to our
previous remark concerning the signs of $\delta\dot\Pi(t)$ and $\dot
P(t)$.  If we now compare cases II and III we see that $\tau$ is
favored for the first choice of constants more than for the second.
Therefore we could interpret $\omega_a$, $\omega_c$, $\Omega_A$ and
$\Omega_C$ as related to the information reaching respectively
$\tau$ and $\sigma$: when $\tau$ gets a larger amount of information he is able to
earn more! However this is clearly only part of the conclusion,
since also in case III the function $\delta\Pi(t)$ is positive, even
if $\Omega_A=\Omega_C=10>1=\omega_a=\omega_c$, so that $\tau$ still increases the value of its portfolio. Moreover, if we
consider the case IV in which $\omega_a$ and $\omega_c$ are even
larger (and much larger than $\Omega_A$ and $\Omega_C$), we see from
our table that $\tau$ does not improve his portfolio as in cases I,
II and III: so what seems to be relevant is $\omega_a-\omega_c$ and
$\Omega_A-\Omega_C$ more than the values of the constants
themselves. However, this is not enough: our numerical results
clearly show that there exists some asymmetry between
$(\omega_a,\omega_c)$ and $(\Omega_A,\Omega_C)$: the $\omega$'s
carry a higher amount of information than that of  the  $\Omega$'s.
This explains also the results in I. The reason for this asymmetry is not
clear, at this moment, and further investigations will be undertaken
soon, even if it could be simply related to the presence of $f(1)$, which only appears in $\R$ and not for $\tau$. A phenomenological law that could display these results, at least when
$\omega_a=\omega_c=:\omega$ and $\Omega_A=\Omega_C=:\Omega$, is
related to a sort of {\em mean value} of $\dot P(t)$: if we put
$\langle\dot P\rangle:=\frac{1}{T}\,\int_0^T\dot P(t)\,dt=\frac{P(T)-P(0)}{T}$,
we see that $\delta\Pi(t)$ has the following expression:
$$
\delta\Pi(t)=\frac{sign(\langle\dot P\rangle)}{g(\omega,\Omega)},
$$
where $g(\omega,\Omega)$ is a certain function which takes its
minimum value for $\omega=\Omega$ and  satisfies the inequality
$g(\omega,\Omega)<g(\Omega,\omega)$ if $\omega>\Omega$.

\section{Conclusions}

In this paper we have continued the analysis of stock markets began in \cite{bag1,bag2}. In particular, the main difference with respect to what we have done in \cite{bag2} consists in the replacement of the simple mechanism which was considered in \cite{bag2} to fix the price of the share with an external field, which is assumed here to be a known function of time. This is useful to get a deeper understanding of the various ingredients of our model, especially in view of further generalizations which should
produce more realistic models,  first introducing several kind of
shares and then looking for a reasonable mechanism which fixes the price
of the shares themselves. It is also necessary to improve the fpl-approximation we have adopted here, in order to avoid the spurious non-conservation of the integrals of motion observed in our numerical results.

It is necessary to stress also that a realistic model of a {\em true} stock market should include also many other mechanisms to take into account other features like short selling. But this surely requires a better starting point than the model discussed here which, however, we believe is  useful to start understanding some basic features and some consequences of our approach.

\vspace{2mm}

We end the paper with few final comments. The first is a physical
one: people with a quantum mechanical background may be surprised that incompatible (i.e. not commuting)  observables appear in the market, and may wonder about  their economical meaning. This problem, however, is only apparent  since all the observables we are interested in form a
\underline {commuting} subset of a larger non-abelian algebra. Therefore they can
be diagonalized simultaneously and a common
orthonormal basis of eigenstates can be indeed obtained as in (\ref{24bis}). The second comment is due to the fact that our
figures show continuous plots while our starting assertion was that
in the stock market we want to describe only discrete quantities play
a role. This is just a matter of interpretation: the continuous
lines that we find solving the differential equations are simply the
ones which interpolate between the real discrete values which are
attained at some particular values of the time which we could
interpret as the {\em time for the transaction}.  Finally, it is clear that another missing aspect of the analysis  proposed so far is a comparison with the already existing methods. However, this comparison will be postponed until we will not produce a sufficiently {\em complete} and {\em realistic} model.

\section*{Acknowledgements}

This work has been financially supported in part by M.U.R.S.T.,
within the  project {\em Problemi Matematici Non Lineari di
Propagazione e Stabilit\`a nei Modelli del Continuo}, coordinated by
Prof. T. Ruggeri.

\vspace{8mm}

 \appendix

\renewcommand{\theequation}{\Alph{section}.\arabic{equation}}

 \section{\hspace{-.7cm}ppendix :  Few results on the number representation}

We discuss here few important facts in
quantum mechanics and second quantization, paying not much attention to mathematical problems arising from the fact that the operators involved are quite often unbounded. More details can be
found, for instance, in \cite{mer,reed} and \cite{brat}, as well as in \cite{bag1,bag2}.

Let $\Hil$ be an Hilbert space and $B(\Hil)$ the set of all the
bounded operators on $\Hil$. $B(\Hil)$ is a C*-algebra, that is an
algebra with involution which is complete under a norm $\|\,.\,\|$
satisfying the so-called C*-property: $\|A^*A\|=\|A\|^2$, for all
$A\in B(\Hil)$. As a matter of fact $B(\Hil)$ is usually seen as a
{\em concrete realization} of an abstract C*-algebra.   Let $\ST$ be
our physical system and $\A$ the set of all the operators useful for
a complete description of $\ST$, which includes the { observables } of $\ST$. For
simplicity it is convenient to assume that  $\A$ is also a
C*-algebra, possibly coinciding with $B(\Hil)$ itself, even if this is not always possible. This aspect,  related with the importance of some unbounded operators within our scheme, will
not be considered here. The description of the time evolution of
$\ST$ is related to a self-adjoint operator $H=H^\dagger$ which is
called {\em the hamiltonian} of $\ST$ and which in standard quantum
mechanics represents  the energy of $\ST$. We will adopt here the
so-called {\em Heisenberg} representation, in which the time
evolution of an observable $X\in\A$ is given by \be
X(t)=e^{iHt}Xe^{-iHt}\label{a1}\en or, equivalently, by the solution
of the differential equation \be
\frac{dX(t)}{dt}=ie^{iHt}[H,X]e^{-iHt}=i[H,X(t)],\label{a2}\en where
$[A,B]:=AB-BA$ is the {\em commutator } between $A$ and $B$. The
time evolution defined in this way is usually a one parameter group
of automorphisms of $\A$.

In our paper a special role is played by the so called {\em
canonical commutation relations } (CCR): we say that a set of
operators $\{a_l,\,a_l^\dagger, l=1,2,\ldots,L\}$ satisfy the CCR if
the following hold:\be [a_l,a_n^\dagger]=\delta_{ln}\1,\hspace{8mm}
[a_l,a_n]=[a_l^\dagger,a_n^\dagger]=0 \label{a3}\en for all
$l,n=1,2,\ldots,L$. Here $\1$ is the identity operator. These
operators, which are widely analyzed in any textbook in quantum
mechanics, see \cite{mer} for instance, are those which are used to
describe $L$ different {\em modes} of bosons. From these operators
we can construct $\hat n_l=a_l^\dagger a_l$ and $\hat N=\sum_{l=1}^L
\hat n_l$ which are both self-adjoint. In particular $\hat n_l$ is
the {\em number operator } for the l-th mode, while $\hat N$ is the
{\em number operator of $\ST$}.

The Hilbert space of our system is constructed as follows: we
introduce the {\em vacuum} of the theory, that is a vector
$\varphi_0$ which is annihilated by all the operators $a_l$:
$a_l\varphi_0=0$ for all $l=1,2,\ldots,L$. Then we act on
$\varphi_0$ with the  operators $a_l^\dagger$ and their powers: \be
\varphi_{n_1,n_2,\ldots,n_L}:=\frac{1}{\sqrt{n_1!\,n_2!\ldots
n_L!}}(a_1^\dagger)^{n_1}(a_2^\dagger)^{n_2}\cdots
(a_L^\dagger)^{n_L}\varphi_0, \label{a4}\en $n_l=0,1,2,\ldots$ for all $l$. These vectors form an
orthonormal set and are eigenstates of both $\hat n_l$ and $\hat N$:
$\hat
n_l\varphi_{n_1,n_2,\ldots,n_L}=n_l\varphi_{n_1,n_2,\ldots,n_L}$ and
$\hat N\varphi_{n_1,n_2,\ldots,n_L}=N\varphi_{n_1,n_2,\ldots,n_L}$,
where $N=\sum_{l=1}^Ln_l$. Moreover using the  CCR we deduce that
$\hat
n_l\left(a_l\varphi_{n_1,n_2,\ldots,n_L}\right)=(n_l-1)(a_l\varphi_{n_1,n_2,\ldots,n_L})$
and $\hat
n_l\left(a_l^\dagger\varphi_{n_1,n_2,\ldots,n_L}\right)=(n_l+1)(a_l^\dagger\varphi_{n_1,n_2,\ldots,n_L})$,
for all $l$. For these reasons the following interpretation is
given: if the $L$ different modes of bosons of $\ST$ are described
by the vector $\varphi_{n_1,n_2,\ldots,n_L}$, this implies that
$n_1$ bosons are in the first mode, $n_2$ in the second mode, and so
on. The operator $\hat n_l$ acts on $\varphi_{n_1,n_2,\ldots,n_L}$
and returns $n_l$, which is exactly the number of bosons in the l-th
mode. The operator $\hat N$ counts the total number of bosons.
Moreover, the operator $a_l$ destroys a boson in the l-th mode,
while $a_l^\dagger$ creates a boson in the same mode. This is why $a_l$ and
$a_l^\dagger$ are usually called the {\em annihilation} and the {\em
creation} operators.

The Hilbert space $\Hil$ is obtained by taking the closure of the linear
span of all these vectors.

 \vspace{2mm}

An operator $Z\in\A$ is a {\em constant of motion} if it commutes
with $H$. Indeed in this case equation (\ref{a2}) implies that $\dot
Z(t)=0$, so that $Z(t)=Z$ for all $t$.

The vector $\varphi_{n_1,n_2,\ldots,n_L}$ in (\ref{a4}) defines a
{\em vector (or number) state } over the algebra $\A$  as
\be\omega_{n_1,n_2,\ldots,n_L}(X)=
\langle\varphi_{n_1,n_2,\ldots,n_L},X\varphi_{n_1,n_2,\ldots,n_L}\rangle,\label{a5}\en
where $\langle\,,\,\rangle$ is the scalar product in the Hilbert space $\Hil$. As we have discussed in \cite{bag1,bag2}, these states
are used to {\em project} from quantum to classical dynamics and to
fix the initial conditions of the market.

\end{document}